\documentclass[conference]{IEEEtran}
\IEEEoverridecommandlockouts
\usepackage{cite}
\usepackage{amsmath,amssymb,amsfonts}
\usepackage{algorithmic}
\usepackage{graphicx}
\usepackage{textcomp}
\usepackage{xcolor}
\usepackage[hyphens]{url}

\usepackage{tikz}
\usetikzlibrary{positioning}
\usetikzlibrary{quantikz}
\usepackage[export]{adjustbox}
\usepackage[font=small,skip=0pt]{caption}

\def\BibTeX{{\rm B\kern-.05em{\sc i\kern-.025em b}\kern-.08em
    T\kern-.1667em\lower.7ex\hbox{E}\kern-.125emX}}

\newcommand{\rev}[1]{\textcolor{black}{#1}}

\makeatletter
\newcommand{\linebreakand}{%
  \end{@IEEEauthorhalign}
  \hfill\mbox{}\par
  \mbox{}\hfill\begin{@IEEEauthorhalign}
}
\makeatother
\makeatletter
\patchcmd{\@maketitle}
  {\addvspace{0.5\baselineskip}\egroup}
  {\addvspace{-1.5\baselineskip}\egroup}
  {}
  {}
\makeatother

\begin{document}

\title{QuTracer: Mitigating Quantum Gate and Measurement Errors by Tracing Subsets of Qubits}
\author{\IEEEauthorblockN{Peiyi Li\IEEEauthorrefmark{1}\thanks{\IEEEauthorrefmark{1} The first two authors contribute equally to this work.}}
\IEEEauthorblockA{\textit{North Carolina State University}\\
Raleigh, USA \\
pli11@ncsu.edu}
\and
\IEEEauthorblockN{Ji Liu\IEEEauthorrefmark{1}}
\IEEEauthorblockA{\textit{Argonne National Laboratory}\\
Lemont, USA \\
ji.liu@anl.gov}
\and
\IEEEauthorblockN{Alvin Gonzales}
\IEEEauthorblockA{\textit{Argonne National Laboratory}\\
Lemont, USA \\
agonzales@anl.gov}
\linebreakand
\IEEEauthorblockN{Zain Hamid Saleem}
\IEEEauthorblockA{\textit{Argonne National Laboratory}\\
Lemont, USA \\
zsaleem@anl.gov}
\and
\IEEEauthorblockN{Huiyang Zhou}
\IEEEauthorblockA{\textit{North Carolina State University}\\
Raleigh, USA \\
hzhou@ncsu.edu}
\and
\IEEEauthorblockN{Paul Hovland}
\IEEEauthorblockA{\textit{Argonne National Laboratory}\\
Lemont, USA \\
hovland@mcs.anl.gov}
}

\maketitle
\thispagestyle{plain}
\pagestyle{plain}


\begin{abstract}

Quantum error mitigation plays a crucial role in the current noisy-intermediate-scale-quantum (NISQ) era. As we advance towards achieving a practical quantum advantage in the near term, error mitigation emerges as an indispensable component. One notable prior work, Jigsaw, demonstrates that measurement crosstalk errors can be effectively mitigated by measuring subsets of qubits. Jigsaw operates by running multiple copies of the original circuit, each time measuring only a subset of qubits. The localized distributions yielded from measurement subsetting suffer from less crosstalk and are then used to update the global distribution, thereby achieving improved output fidelity.

Inspired by the idea of measurement subsetting, we propose QuTracer, a framework designed to mitigate both gate and measurement errors in subsets of qubits by tracing the states of qubit subsets throughout the computational process. 
In order to achieve this goal, we introduce a technique, qubit subsetting Pauli checks (QSPC), which utilizes circuit cutting and \rev{Pauli Check Sandwiching (PCS)} to trace the qubit subsets distribution to mitigate errors. 
The QuTracer framework can be applied to various algorithms including, but not limited to, VQE, QAOA, quantum arithmetic circuits, QPE, and Hamiltonian simulations. In our experiments, we perform both noisy simulations and real device experiments to demonstrate that QuTracer is scalable and significantly outperforms the state-of-the-art approaches.

\end{abstract}

\setcounter{page}{1}

\section{Introduction}
Quantum computing is rapidly emerging as a transformative technology, offering great potential for chemistry simulations~\cite{lanyon2010towards}, combinatorial optimization~\cite{farhi2014quantum}, machine learning~\cite{schuld2015introduction}, and other domains~\cite{Buhrman_2014PosBasedQuantCryptoImp&Const}.  Ideal quantum computers with fully fault-tolerant error correction codes \cite{Aharonov_2008FaultTolQuantCompWithConstErrRate, Kitaev_1997QuantCompAlgoAndErrCorr} remain distant, and we currently find ourselves in the Noisy Intermediate Scale Quantum (NISQ) era~\cite{preskill2018quantum}, characterized by quantum computers comprising of tens to thousands of noisy qubits and limited connectivity.

In the NISQ era, quantum error mitigation has emerged as a promising strategy to deal with errors arising during quantum computation. Instead of fully correcting the errors, we may mitigate these errors to an acceptable level. Various approaches have been proposed to this end, including Zero Noise Extrapolation~\cite{temme2017error, Giurgica_2020DigitZNEFroQEM}, Clifford Data Regression~\cite{Czarnik_2021errormitigationWithCliffQCData}, Virtual Distillation~\cite{huggins2021virtual, Koczor_2021ExpErrSuppForNearTermQuantDevices}, Symmetry Verification~\cite{BonetMonroig_2018LowCostErrMitigBySymmVer, McArdle_2019ErrMitigDigitalQuantSim, McClean_2020DecodingQuantErrorsWithSubspaceExpansions, Cai_2021QuantErrMitigUsingSymmExpansion, tsubouchi_2023VirtualQuantErrDetect}, Pauli Check Sandwiching (PCS)~\cite{Debroy_2020ExtendFlagGadgetsForLOCircVer, gonzales2023pcs, van2023singlesidePCS}, and measurement subsetting~\cite{Das_2021JigSawSubsetting, Dangwal_2023varsawAppTMeasEMForVQA}.

Previous work in measurement subsetting~\cite{Das_2021JigSawSubsetting, Dangwal_2023varsawAppTMeasEMForVQA} made the observation that measuring a subset of qubits leads to lower measurement errors than measuring all the qubits by reducing the measurement crosstalk. The more accurate local distributions from subset measurements can then be used to improve the measurement results of all the qubits, i.e., the global distribution. 
Inspired by measurement subsetting, we propose the following hypothesis: if we can mitigate the noise of a subset of qubits throughout the entire circuit execution and achieve high-fidelity results for these subsets of qubits, can we significantly improve the overall fidelity? In this paper, we develop a novel qubit subsetting framework, QuTracer, to validate our hypothesis. QuTracer continuously tracks the state of qubit subsets throughout the computational process and mitigates errors along the way. The high-fidelity local distributions can be obtained and used to refine the noisy global distribution. Compared to the previous work on measurement subsetting, QuTracer effectively mitigates both gate and measurement errors. Mitigating noise in a subset of qubits is also much more effective and less costly than directly mitigating the noise on all the qubits.

QuTracer consists of multiple novel ideas. First, we repurpose the circuit cutting technique to track the state of qubit subsets, analogous to watchpoints during program execution. Second, we propose qubit subsetting Pauli check (QSPC)  to mitigate the errors on qubit subsets. Our proposed implementation of QSPC has low circuit cost. For example, adding single-qubit subset Pauli checks only requires insertion of single-qubit gates. Third, we propose a multi-layer qubit subsetting approach designed to mitigate errors in circuits requiring multiple layers of Pauli checks.
Fourth, we propose a series of novel optimizations specifically tailored for the process of qubit subsetting, offering improvements in efficiency and accuracy. 

Our noisy simulator and real-device experimental results confirm the effectiveness of our proposed QuTracer framework. 

Our contributions can be summarized as follows:
\begin{itemize}
\item We introduce QuTracer, a qubit subsetting framework that monitors the distribution of a subset of qubits to mitigate errors. It mitigates both gate errors and measurement crosstalk errors, thereby surpassing the capability of measurement subsetting.
\item We repurpose circuit cutting to track the quantum states during the circuit execution. This is analogous to enabling watchpoints when debugging classical programs.
\item We propose qubit subsetting Pauli Checks (QSPC) to mitigate the errors on qubit subsets. QSPC has a lower cost than the existing Pauli check error detection schemes.
\item We propose and incorporate multiple optimizations for qubit subsetting. These optimizations include state preparation reduction, localized gate simulation, gate bypassing, state traceback, false dependency removal, and qubit remapping.
\item We demonstrate the scalability and effectiveness of our approach through rigorous experimentation. Our experimental results on both noisy simulators and real quantum devices show that QuTracer significantly outperforms the state-of-the-art approaches.
\end{itemize}

This paper is organized as follows. In Section~\ref{sec: background}, we review related works and discuss the motivation of qubit subsetting. In Section~\ref{sec:qubit subsetting iQFT}, we use the inverse quantum Fourier transform circuit as an example to illustrate the QuTracer framework.
In Section~\ref{sec:qubit subsetting checks}, we present the theory and the design of qubit subsetting Pauli checks. In Section~\ref{sec:qutracer_framework}, we present the design of the QuTracer framework and discuss the relevant optimizations. In Section~\ref{sec:methodology}, we provide the experimental setup. In Section~\ref{sec:evaluation}, we discuss both the noisy simulation and the real-device results. Finally, Section~\ref{sec:conclusion} concludes the paper.

\section{Background and Motivation}
\label{sec: background}

\subsection{Measurement Error Mitigation via Measurement Subsetting}
Measurement subsetting, i.e., the JigSaw protocol, aims to reduce the effects of measurement errors \cite{Das_2021JigSawSubsetting}. The protocol splits the experiment shots into a set containing all the end measurements and a set containing only partial measurements. The circuits with partial measurements are equivalent to the complete circuit, except that some of the measurement operators at the end are removed. The circuit with all measurements generates a noisy global distribution. The partial measurement circuits generate local distributions that have high local fidelities due to the reduced measurement errors. Then, the local distributions refine the global distribution through a Bayesian recombination method that uses local information to update the global distribution. VarSaw is an improved version of JigSaw intended for variational quantum algorithms (VQA) \cite{Dangwal_2023varsawAppTMeasEMForVQA}. It recognizes that various time and spatial redundancies exist in measurement subsetting for VQA circuits. By eliminating such redundancy, VarSaw can achieve better efficiency.

\textbf{Takeaway:} While Jigsaw and Varsaw target measurement errors, gate errors remain unmitigated. The effectiveness of measurement subsetting is limited for circuits with a high circuit depth, leading to a high gate error rate. Therefore, a subsetting technique that can also effectively mitigate gate errors is desired.

\subsection{Circuit Cutting}
Quantum circuit cutting decomposes a payload circuit into smaller fragments such that they can run on smaller quantum devices \cite{Peng_2020CircuitCutting, Ayral_2020QuantDivide&Comp:HDandNoisSim, Perlin_2021QuantCircCutWithMLTomog, Ayral_2021QuantDiv&Comp:ExpTheEffOfDiffNoisSour}. Circuit cutting works by measuring a complete basis set on the left side of the cut and preparing the corresponding states on the right side of the cut. To see how this can be done when cutting one qubit, we can fragment an arbitrary quantum state $\rho$ as
\begin{align}
    \rho=\frac{1}{2}\sum_{M\in  \mathcal{B}}M\otimes\text{tr}_j(M_j\rho),
\label{eq:rho_decompose}
\end{align}
where $\mathcal{B}$ represents the basis set of $2\times 2$ Pauli matrices, $\text{tr}_j$ is the partial trace over qubit $j$,
and $M_j$ represents an operator that acts with Pauli operator $M$ on qubit $j$ and acts with identity $I$ on other qubits. By expanding $M$ in its spectral decomposition, we can decompose $\rho$ into a sum of channels that contains measurements $M$ on qubit $j$ followed by the preparation of quantum states which are the eigenstates of $M$. When cutting multiple qubits, \rev{the number of Pauli basis scales exponentially $O(4^n)$ with the number of the cutting qubits $n$.} 

Measuring the probability distribution of qubits at the cut provides complete information about the quantum state at that position. In conventional circuit cutting, this state information is used to reconstruct the state after the circuit cutting. However, we make a key observation that the state information at the cut can also be leveraged to monitor the execution of the program and to facilitate error mitigation.

\textbf{Takeaway:} Circuit cutting, rather than being solely used for dividing a circuit into multiple subcircuits, can be repurposed to monitor the quantum states throughout the execution of a circuit. 
\begin{figure}[htbp]
  \centering
    \begin{adjustbox}{width=\linewidth}
    \begin{quantikz}[]
    \lstick{\ket{0}} & \gate{H} & \ctrl{1} & \qw & \qw  & \ctrl{1} & \gate{H} & \meter{} \\
    \lstick[wires=3]{$\rho$}& \qw & \gate[3]{C_L} & \gate[3]{U} & \gate[3]{\varepsilon} & \gate[3]{C_R} &  \qw & \rstick[wires=3]{$\rho_{out}$} \qw\\
    & \qw  & \qw  & \qw & \qw & \qw & \qw & \qw \\
    & \qw  & \qw  & \qw & \qw & \qw & \qw & \qw & 
    \end{quantikz} 
    \end{adjustbox}
    
    \begin{adjustbox}{width=\linewidth}
    \begin{quantikz}[]
    \lstick{\ket{0}} & \gate{H} & \qw & \ctrl{1} & \qw  & \ctrl{1} & \gate{H} & \meter{} \\
    \lstick[wires=3]{$\rho$}& \qw & \gate[3]{U} & \gate[3]{C_R^{\dagger}} & \gate[3]{\varepsilon} & \gate[3]{C_R} &  \qw & \rstick[wires=3]{$\rho_{out}$} \qw\\
    & \qw  & \qw  & \qw & \qw & \qw & \qw & \qw \\
    & \qw  & \qw  & \qw & \qw & \qw & \qw & \qw & 
    \end{quantikz} 
    \end{adjustbox}
\caption{General idea of Pauli Check Sandwiching (PCS). $\varepsilon$ is the noise map due to $U$. The two circuits are equivalent as a result of Eq.~\eqref{eq:PCSMultiCond}, and the sandwiched $\varepsilon$ can be seen as a transformed noise map. \rev{The gates that prepare the input state $\rho$ are not shown.}}
\label{fig:PCSscheme}
\end{figure}
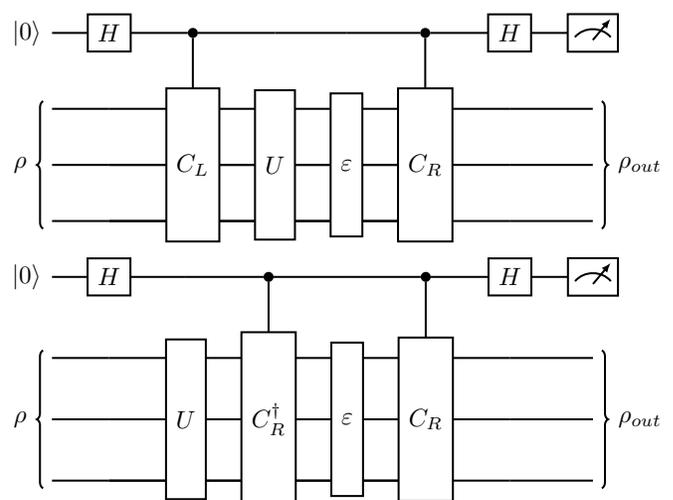

\begin{figure*}
\centering
{\includegraphics[width=\linewidth]{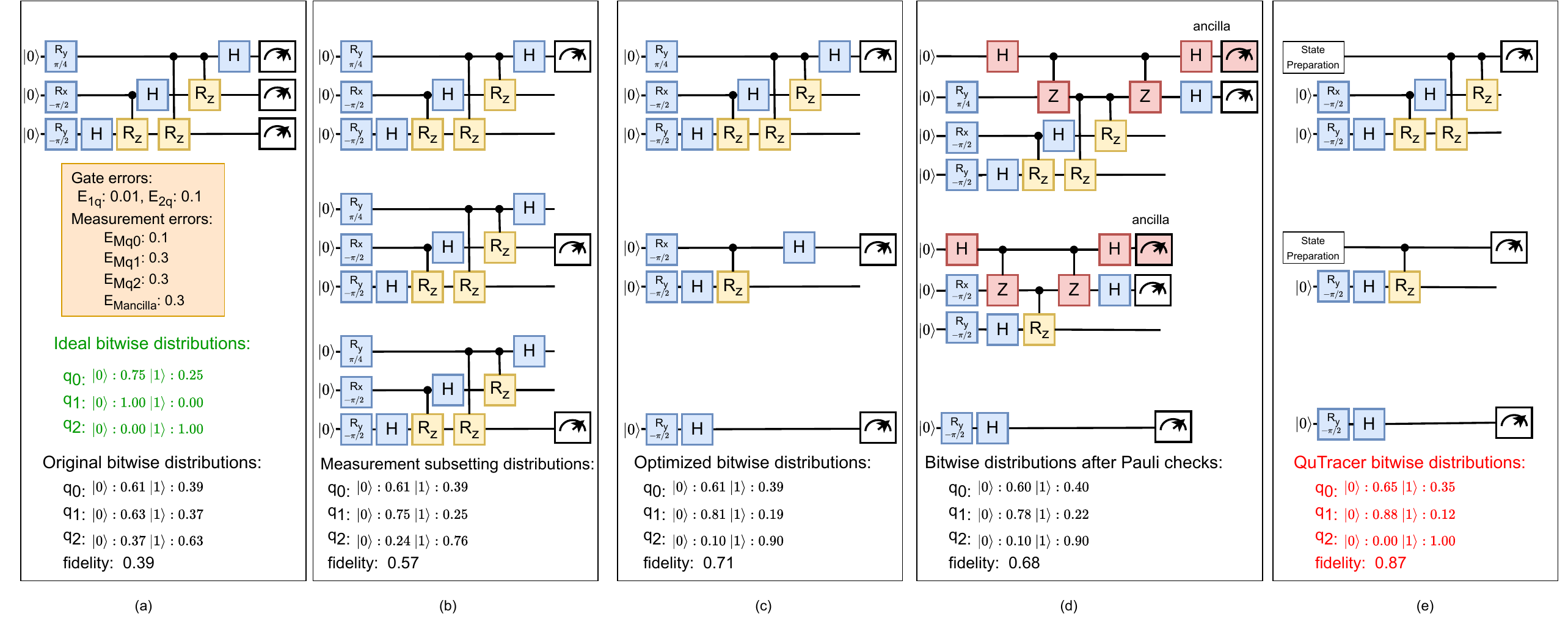}
}\hfil
\caption{QuTracer optimized circuits and the corresponding bitwise distributions (i.e., local distribution with subset size of 1) and output Hellinger fidelity (i.e., fidelity of global distribution). (a) The original iQFT circuit that generates the noisy global distribution; (b) Circuit copies with a measurement subsetting size of one; (c) Optimized circuit copies of (b) by removing the gates that the measurement subset has no dependence upon; (d) Circuit copies with PCS to mitigate gate errors; (e) The QuTracer optimized circuit copies of (d).}
\label{fig:iqft_example}
\end{figure*}

\subsection{Pauli Check Sandwiching (PCS) and SQEM framework}
PCS~\cite{gonzales2023pcs} is a technique designed to detect errors in quantum circuits. This method involves applying a set of checks that have known transformations with the payload circuit to mitigate the errors. 

The $n$-qubit Pauli group $P_n$ is defined as
\begin{equation}
    P_n = \{I, X, Y, Z\}^{\otimes n} \times \{\pm1, \pm i\}.
\end{equation}
Let $U$ denote the payload unitary circuit. Pauli checks are chosen such that the following constraint is satisfied
\begin{align}\label{eq:PCSMultiCond}
    C_{R}UC_{L}= U
\end{align}
and typically, $C_{L}$ and $C_{R}$ are elements of the Pauli group.
For a single pair of checks, the PCS scheme applies the circuit shown in Fig.~\ref{fig:PCSscheme} and post-selects on the zero outcome of a Z-basis measurement of the ancilla qubit. 
Note that the right check along with the post-selection can be recognized as the Hadamard test often used in quantum error correction \cite{nielsen2011quantumCompAndQuantInfo}. The single layer scheme and its equivalent form are depicted in Fig.~\ref{fig:PCSscheme}. As long as the error operator anti-commutes with $C_R$, it can be effectively eliminated as a result of post-selection. 

In PCS and other error mitigation techniques, the error mitigation protocol introduces extra errors via additional gates (e.g., the Hadamard test gates) and ancilla qubits. This extra noise degrades the performance of the error mitigation schemes. Simulated Quantum Error Mitigation (SQEM) was developed to minimize the added noise \cite{liu_2022classicalSimsAsQEMviaCircCut} by leveraging circuit cutting. In SQEM, circuit cutting is performed so that the payload subcircuit becomes separated from the quantum error mitigation subcircuit. The payload circuit fragment runs on the quantum hardware and the quantum error mitigation fragments are simulated on a classical computer. The results of the fragments are recombined via classical post-processing.

Although SQEM addresses the issue of additional noise from mitigation circuits, it inherits the scalability challenges of circuit cutting. When mitigating multi-layer VQE or QAOA circuits, SQEM requires cutting the error mitigation circuit across layers, leading to an overhead that scales exponentially with the number of layers. 

\textbf{Takeaway:} PCS is a promising mitigation technique but the extra noise due to its additional gates and ancillas limits its effectiveness. Although the SQEM framework reduces this extra noise of PCS and can be used for shallow VQE circuits, there is a need for a more generalized framework capable of handling multi-layer, complex circuits. The crucial part is to address the scalability resulting from circuit cutting.

\section{Motivating Example: Inverse QFT}
\label{sec:qubit subsetting iQFT}
In this section, we use the inverse quantum Fourier transform (iQFT) circuit as an example to show the key idea behind the QuTracer framework. QFT and iQFT are basic building blocks in many quantum algorithms including Shor's algorithm~\cite{shor1994algorithms}, quantum adder~\cite{draper2000adder}, quantum multiplexer~\cite{ruiz2017quantum_multilexer}, quantum phase estimation~\cite{kitaev1995QPE}, and HHL algorithm~\cite{harrow2009quantum_HHL}. A three-qubit iQFT circuit is shown in Fig.~\ref{fig:iqft_example}(a). When running on a noisy simulator that incorporates both simulated gate and measurement errors, the Hellinger fidelity of the output state is 0.39 due to the noise in both circuit execution and qubit measurements. \rev{The measurement error vary from 0.1 to 0.3.}

The key ideas behind our proposed qubit subsetting are that (a) it optimizes the circuit copies generated for qubit subsets, and (b) it enables error mitigation during the computation of these qubit subsets. 
The circuits in Fig.~\ref{fig:iqft_example}(b) employ measurement subsetting, i.e., Jigsaw, with a subset size of one, resulting in three separate circuit copies for this three-qubit iQFT. \rev{Jigsaw maps the qubit subset to qubits with lower measurement errors, and the resulting output fidelity becomes 0.57 after combining the more accurate local distributions.} In comparison, QuTracer further optimizes these circuits. First, the gates that the measurement operator has no dependency on can be removed. Second, as we only measure the probability distribution on the Z basis, the Rz gates, which only induce a phase change, can also be eliminated. The optimized circuits are shown in Fig.~\ref{fig:iqft_example}(c). Note that this optimization opportunity is unique and arises specifically when the focus is on the distribution of a subset of qubits. As the optimized circuits contain fewer gates, local distributions can be further improved, and the fidelity of the refined global state becomes 0.71. 

After optimizing the circuit, we apply error mitigation schemes. Fig.~\ref{fig:iqft_example}(d) shows the circuits after applying PCS. Here, Pauli-Z checking is used as the Z gate on the control qubit commutes with the control-Z gate to be protected. By introducing an ancilla qubit and the Pauli-Z checking circuit, errors that anti-commute with the Pauli-Z operator (e.g., Pauli-X and Pauli-Y) can be detected. The ancilla qubit and the extra gates are marked in red in the figure. However, these extra operations themselves introduce noise, and there is no guarantee that the noise reduction would outweigh the induced errors due to the extra gates. In our experiment, this added mitigation actually showed worse local distributions than Fig.~\ref{fig:iqft_example}(c), and the output fidelity becomes 0.68. 
To overcome the noise due to the mitigation circuits, we propose to implement these mitigation operations ``virtually." As elaborated in the following section, we develop a qubit subsetting Pauli check (QSPC) technique, which transforms these operations into a collection of lower-cost circuits that introduce only minimal noise. Consequently, the circuits depicted in Fig.~\ref{fig:iqft_example}(d) are transformed into the more efficient ones shown in Fig.~\ref{fig:iqft_example}(e), which involve state preparation and measurements, but only introduce additional single-qubit gates, which have much lower error rates than two-qubit ones. With QSPC, the local distributions become more accurate as a result of mitigating both gate and measurement errors. The fidelity of the refined global distribution becomes 0.87, a 53\% improvement over Jigsaw.

\section{Qubit subsetting Pauli Checks (QSPC)}
\label{sec:qubit subsetting checks}

In this section, we present Qubit Subsetting Pauli Checks (QSPC), a novel approach specifically designed to check a subset of qubits. 

\subsection{Intuition}
QSPC incorporates two key elements. First, it employs PCS protocol to mitigate gate errors. 
Second, it virtualizes the PCS circuits by transforming them into an ensemble of state preparation and measurements. This is achieved by leveraging the circuit-cutting idea. As a side benefit, some quantum gates can be replaced with classical computation if they depend solely on the state at a cut. Moreover, as the post-selected output states can be prepared directly from the cut, there is no need to measure the ancilla qubits, thus QSPC is immune to measurement errors on ancilla qubits. Virtualization of the PCS circuit also offers a unique advantage: it integrates the measurement errors from the original circuit into the error channel protected by the Pauli checks. This results in the virtual Pauli checks effectively mitigating both gate and measurement errors, a significant improvement over the original PCS that only addresses gate errors. 

\subsection{Theory}
\label{subsec:theory_QSPC}
The circuit implementation of the PCS protocol shown in Fig.~\ref{fig:PCSscheme}, uses a pair of n-qubit gates $\left\{C_L, C_R\right\}$ to check the n-qubit circuit $U$. The requirement is $C_RUC_L = U$. With the error channel being $\epsilon(\rho) = \sum_iE_i \rho E_i^{\dagger}$, the post-selected output state of the PCS protocol is~\cite{gonzales2023pcs}:

\begin{equation}
\resizebox{.9\hsize}{!}{
$\rho_{out}= \frac{\sum_i\left[\left(C_R E_i C_R^{\dagger} + E_i\right) U \rho U^{\dagger}\left(C_R E_i^{\dagger} C_R^{\dagger} + E_i^{\dagger}\right) \right]}{\operatorname{tr}\left(\sum_i\left[\left(C_R E_i C_R^{\dagger} + E_i\right) U \rho U^{\dagger}\left(C_R E_i^{\dagger} C_R^{\dagger} + E_i^{\dagger}\right) \right]\right)}$}
\label{eq:post_output}
\end{equation}

As long as the error operator $E_i$ anticommutes with $C_R$,  $C_RE_iC_R^{\dagger} + E_i = 0$ and the error can be eliminated as a result of post-selection. 
When there is no error, i.e., $E_i=I$, 
$\rho_{out}=U \rho U^{\dagger}$, meaning that the execution of the gate $U$ is noise free. 
In QSPC, we use Pauli-Z checks to protect a subset of qubits, the subset size of qubits is set to 1, 
i.e., $C_R=C_L=Z_j$, where $Z_j$ represents an operator that acts with Pauli operator Z on qubit $j$ and acts with identity $I$ on other qubits, and qubit $j$ is the qubit that we want to protect.
Pauli-Z checks capture bitflip errors on the protected qubit as X anti-commutes with Z. 
The discussion of protecting more than one qubit simultaneously, i.e., setting the qubit subset size to be larger than 1 will be introduced in section~\ref{subsec:subset_size}.

Generating the output state $\rho_{out}$ shown in Eq.~(\ref{eq:post_output}) requires ancilla qubit measurement and post-selection, which introduce extra noise. To address the issue, we make the observation that it is sufficient to obtain the expectation value of $\rho_{out}$ with respect to an observable $O$, $\left \langle O\right \rangle =\operatorname{tr}(\rho_{out}O)$, instead of the exact state $\rho_{out}$. There are two reasons for this. Firstly, in many algorithms, including variational ones, the expectation value of an observable is needed instead of the output distribution. Secondly, measuring the probability of a classical output is actually equivalent to measuring the corresponding observables. The benefit of using $\operatorname{tr}(\rho_{out}O)$ is that it can be computed virtually by leveraging the circuit cutting technique. 
To compute $\operatorname{tr}(\rho_{out}O)$, we first derive the following relationship: since $C_RUC_L = U$, we have $C_R^{\dagger} U = UC_L$ and $U^{\dagger} C_R = C_L^{\dagger}U^{\dagger}$. Then, we use Eq.~(\ref{eq:post_output}) to compute the numerator and denominator of $\operatorname{tr}(\rho_{out}O)$ separately. The numerator of $\operatorname{tr}(\rho_{out}O)$ can be computed from the following four terms: 
\begin{align}
\label{eq:expandObservable1}
& \operatorname{tr}(\Sigma_i E_i U \rho U^{\dagger} E_i^{\dagger} O) \\
\label{eq:expandObservable2}
& \operatorname{tr}(\Sigma_i E_i U C_L \rho U^{\dagger} E_i^{\dagger} O C_R )\\
\label{eq:expandObservable3}
& \operatorname{tr}(\Sigma_i E_i U \rho C_L^{\dagger} U^{\dagger} E_i ^{\dagger}C_R^{\dagger} O )\\
\label{eq:expandObservable4}
& \operatorname{tr}(\Sigma_i E_i U C_L \rho C_L^{\dagger} U^{\dagger} E_i^{\dagger} C_R^{\dagger} O C_R ).
\end{align}

\begin{figure*}
\centering
{\includegraphics[width=\linewidth]{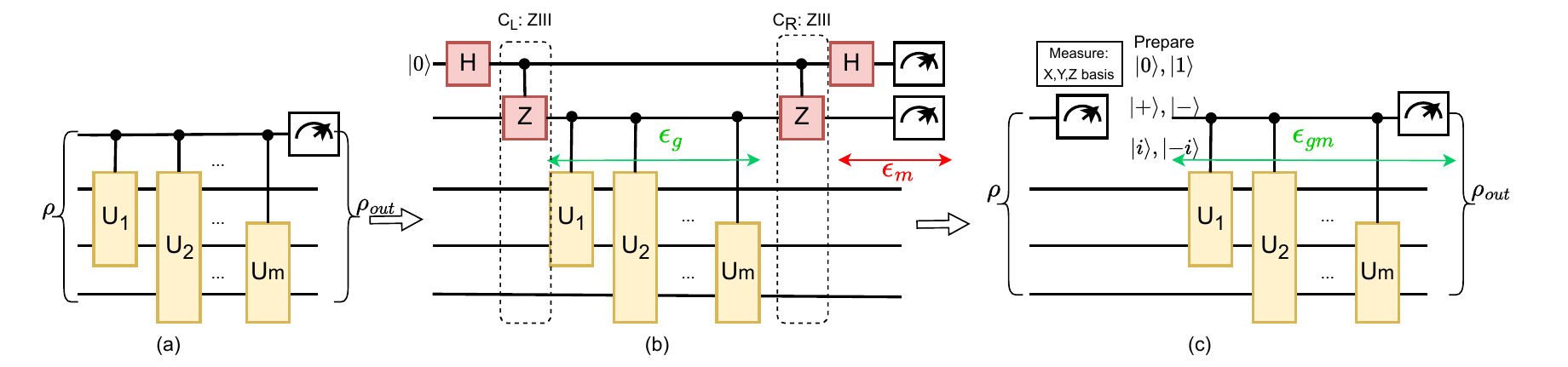}
}\hfil
\caption{Qubit subsetting Pauli check for a single qubit. (a) Circuit with one qubit to measure as a result of measurement subsetting; (b) Circuit with Pauli-Check Sandwiching; (c) Circuit with our proposed qubit subsetting Pauli checking. $\epsilon_g$ \rev{denotes} the error channel that consists of gate errors. $\epsilon_m$ denotes the measurement error channel. $\epsilon_{gm}$ denotes the error channel that consists of both gate and measurement errors. \rev{The sequence of gates $U_{1-m}$ sharing the same control qubit exists widely in algorithms such as QFT and QPE. We use this pattern to illustrate QuTracer's ability to protect a series of operations commuting on a subset of qubits. }}
\label{fig:subset_check_example}
\end{figure*}

These four terms can be obtained by preparing $\rho, C_L\rho , \rho C_L^{\dagger}, $ and $C_L \rho C_L^{\dagger}$, passing them through a quantum gate $U$ followed by a noisy quantum channel $\epsilon(\rho)$, and measuring four observables $O, O C_R , C_R^{\dagger}O$, and $C_R^{\dagger}OC_R$, respectively.
This process consists solely of state preparation and measurement and do not involve ancilla qubits and controlled unitary gates which are required in the PCS circuits. Therefore, the errors typically associated with the QSPC checking circuit can be significantly reduced. 
We will detail the procedure for calculating Term (\ref{eq:expandObservable2}); other terms can be calculated by the same procedure.

For Term (\ref{eq:expandObservable2}), we need to first prepare $C_L \rho $. But $C_L \rho $ is not a quantum state that can be directly prepared. 
However, based on Eq.~(\ref{eq:rho_decompose}), $C_L \rho $ can be decomposed as Eq.~(\ref{eq:CL_rho_decompose}), which is a linear combination of channels that contains measurements followed by state preparation. This allows us to use a similar process as the wire cutting in quantum circuits to obtain $C_L \rho $. 
\begin{align}
 C_L \rho=\frac{1}{2}\sum_{M\in \mathcal{B}} Z M\otimes\text{tr}_j(M_j\rho),
\label{eq:CL_rho_decompose}
\end{align}

Therefore, the preparation of $C_L \rho $ transforms into setting up a series of circuits involving measurements followed by state preparation. In each circuit, the measurement operator $M_j$ is applied to the state $\rho$, followed by preparing the eigenstates of $ZM$ on the $j$th qubit. Then based on Term (\ref{eq:expandObservable2}), all these circuits go through the quantum gate $U$ and noisy quantum channel $\epsilon(\rho)$, the observable $O C_R$ is measured to obtain the results. 
Fig.~\ref{fig:subset_check_example}(c) gives an example of setting up the circuits to compute Term (\ref{eq:expandObservable2}) when the checking qubit is the first qubit. 
The results from all the circuits are combined linearly to derive the outcome for Term (\ref{eq:expandObservable2}).

We can calculate all the terms for the numerator of $\operatorname{tr}(\rho_{out}O)$ by the above procedure. The calculation for the denominator of $\operatorname{tr}(\rho_{out}O)$ is in the same manner with the observable $O$ being the identity $I$ for all the qubits. 
A maximum of 18 distinct circuits is sufficient to calculate $\operatorname{tr}(\rho_{out} O)$.
In some scenarios, we care about the information of the protected qubit of $\rho_{out}$ in all three measurement basis $X$, $Y$, and $Z$. Then in the worst-case scenario, we need to prepare 30 different circuits to calculate $\operatorname{tr}(\rho_{out} X_j)$, $\operatorname{tr}(\rho_{out} Y_j)$, and $\operatorname{tr}(\rho_{out} Z_j)$ by reusing results from some circuits.

\subsection{QSPC vs. SQEM}
\label{subsec:qspc_vs_sqem}
In our approach, the transformation of the PCS circuit into an ensemble of state preparations and measurements is akin to the process of cutting and simulating Pauli check circuits, as discussed in the SQEM framework~\cite{liu_2022classicalSimsAsQEMviaCircCut}. However, the key distinction between our Qubit Subsetting Pauli Checks (QSPC) and the SQEM approach lies in the state reconstruction process. Circuit cutting necessitates preparing and measuring on all bases for complete state reconstruction. \rev{For a cut fragment with $m$ measurement locations and $n$ state preparation locations, the standard circuit cutting requires $3^m \times 4^n$ circuit copies, as detailed in~\cite{Perlin_2021QuantCircCutWithMLTomog}. Since there are two measurement locations and one state preparation location when cutting the circuit in Fig.~\ref{fig:subset_check_example}(b), SQEM requires 36 circuit copies. In contrast, QSPC directly calculates the necessary state preparations and measurement bases, thus significantly reducing the number of required state preparations and measurements. As mentioned earlier, the calculation requires 18 circuits. Even in the worst case scenario, QSPC surpasses SQEM, requiring 30 circuits.} In the experimental section (Section~\ref{sec:evaluation}), we will present the overhead for various benchmarks and demonstrate the substantial reduction in circuit cost achieved through our method.

\begin{figure*}
\centering
{\includegraphics[width=\linewidth]{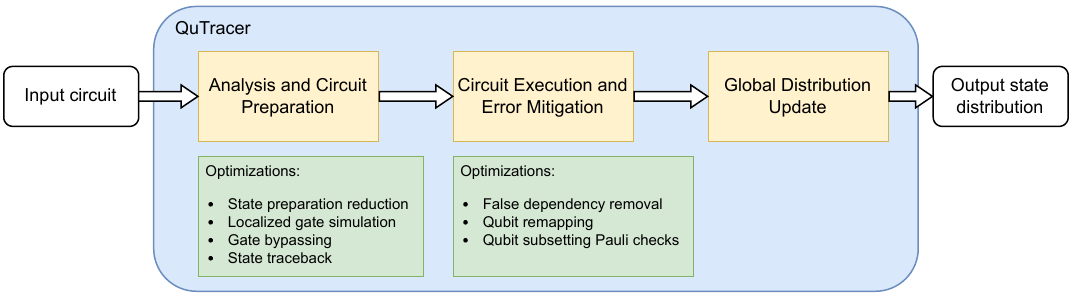}
}\hfil
\caption{QuTracer workflow and the optimizations}
\label{fig:Qutracer}
\end{figure*}

\subsection{Measurement Error Mitigation}
QSPC offers a unique advantage in that it enables the mitigation of both gate and measurement errors. In the original PCS scheme shown in Fig.~\ref{fig:subset_check_example}(b), the error channel $\epsilon_g$ consists of gate errors that occur between the two checks. The measurement errors denoted by $\epsilon_m$ are not mitigated.  However, with QSPC’s “virtual” implementation of checks, the final measurements on the qubits are also sandwiched between these checks. As we prepare states and measure observables in QSPC, the resultant error channel, denoted $\epsilon_{gm}$, consists of both gate and measurement errors. The post-selected output state that we reconstructed is the state that mitigates both gate and measurement errors. This is particularly effective when measurement errors can be viewed as products of bit-flip Pauli X errors. Since these measurement errors anticommute with Pauli Z checks, they can be effectively mitigated using the QSPC method. 

In summary, QSPC virtualizes the single-qubit PCS circuit in Fig.~\ref{fig:subset_check_example}(b) and converts it to an ensemble of circuits in Fig.~\ref{fig:subset_check_example}(c) that only contains additional single-qubit gates, thus eliminating the use of ancilla qubits and reducing the noise in these checking circuits. In addition, the ensemble of circuits mitigates the measurement errors. In the next section, we will discuss how to incorporate QSPC in our error mitigation framework to protect a general quantum circuit.

\section{QuTracer Framework}
\label{sec:qutracer_framework}
\subsection{Framework}\label{subsec:qspcFramework}

Our QuTracer framework involves continual tracking of the state of qubit subsets throughout their computational process. 

In order to track the state during circuit execution, we re-purpose the circuit-cutting technique: for a cut, we actually measure the distribution at the cutting point and prepare the necessary states accordingly. We can insert multiple cut points on a subset of qubits and measure at these points to track their state at each cut point. In other words, the purpose of these cuts is not to separate a circuit but to track the quantum states to ensure the correct execution of the circuit. This is analogous to creating watchpoints when debugging classical programs. Note that having watchpoints on many qubits incurs exceedingly high overhead. On the other hand, watchpoints are a good fit when used on a small subset of qubits as the overhead would be much more manageable.  
By measuring at the ``quantum watchpoints" and error-mitigating the state of a subset of qubits, we can achieve high-fidelity qubit subset distributions.

The workflow of our approach is as follows. Given a quantum circuit, it is first executed to produce the global distribution. Then, QuTracer performs qubit subsetting to produce high-fidelity local distributions and refines the global one. At a high level, QuTracer takes three distinct steps shown in Fig.~\ref{fig:Qutracer}:
\begin{itemize}
    \item \textbf{Analysis and Circuit Preparation:} For a circuit with a subset of qubits of interest, analyze the circuit to determine the cut locations where the measurement of the qubit subset states is needed. To measure the states of the qubit subset, multiple state preparation and measurement circuits will be generated following the circuit cutting protocol.
    \item \textbf{Circuit Execution and Error Mitigation:} Execute the circuits and subsequently update the qubit subset states. During this step, error mitigation techniques are strategically employed to ensure an accurate estimation of the subset state.
    \item \textbf{Global Distribution Update:} Refine the global distribution based on the qubit subset states. The global distribution is updated using the same Bayesian recombination algorithms utilized by SQEM~\cite{liu_2022classicalSimsAsQEMviaCircCut}.
\end{itemize}

Within each step, we propose further optimizations as listed in Fig.~\ref{fig:Qutracer}. To facilitate a clear and focused examination of the concepts and optimizations, we limit the size of the qubit subset to one in the following discussion.

\subsection{Single-layer Qubit Subsetting}\label{subsec:QSPCsingleLayer}
\begin{figure*}
\centering
{\includegraphics[width=\linewidth]{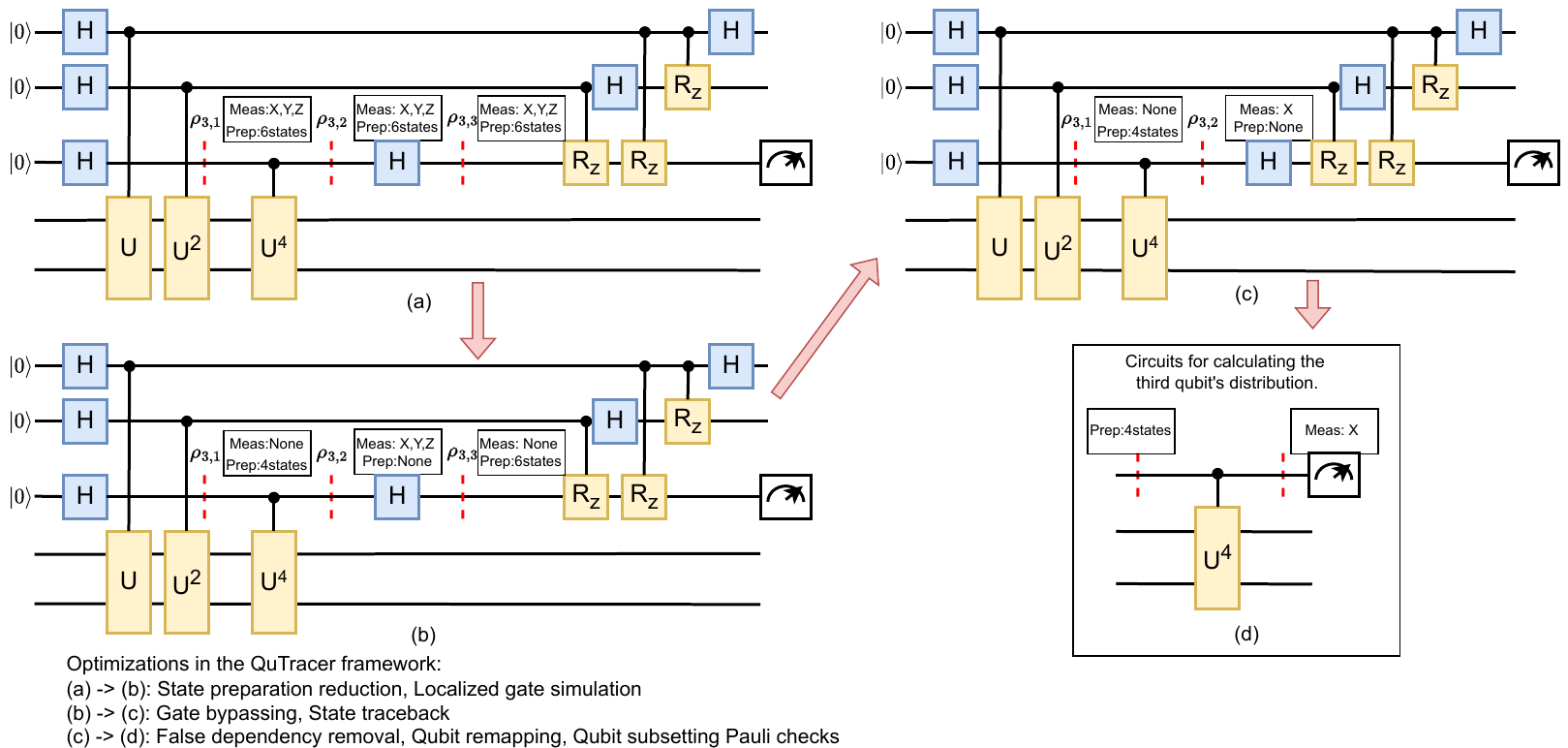}
}\hfil
\caption{QuTracer framework for Quantum Phase Estimation (QPE)}
\label{fig:qpe_example}
\end{figure*}

In this subsection, we use Quantum Phase Estimation (QPE) as an example to illustrate how its execution result can be refined by the QuTracer framework. Fig.~\ref{fig:qpe_example}(a) depicts a quantum phase estimation circuit featuring three ancilla qubits. As only these ancilla qubits are to be measured, qubit subsetting is specifically applied to these three qubits by obtaining the accurate distribution of one ancilla qubit at a time. Here, we focus on the third qubit. As shown in Fig.~\ref{fig:qpe_example}(a), the circuit analyzer strategically inserts three circuit cut points. The criteria for choosing cut points is to divide the gate operations into sets of commuting operations. The reason is that a sequence of commuting operations can be efficiently checked by our Qubit Subsetting Pauli Checks (QSPC) approach. Initially, each cut point requires measurement in three bases and preparation in six states. However, many state preparations and measurements can be optimized. The \textbf{Analysis and Circuit Preparation} step incorporates the following optimizations:

\textbf{State preparation reduction:} In Fig.~\ref{fig:qpe_example}(a), we prepare six states $\ket{0}$, $\ket{1}$, $\ket{+}$, $\ket{-}$, $\ket{i}$ and $\ket{-i}$ at the cut point $\rho_{3,1}$ in order to compute $\rho_{3,2}$. The expectation value of states $\ket{-}$ and $\ket{-i}$ can be derived with classical post-processing based on the expectation value of the other four states $\ket{0}$, $\ket{1}$, $\ket{+}$, and $\ket{i}$. Therefore, the number of state preparations in the cut point $\rho_{3,1}$ can be reduced to four.

\textbf{Localized gate simulation:} Since we restrict the size of the subsetting, the localized gates that operate only on the subset of qubits can be efficiently simulated using classical computers. In our example, we track the state on the third qubit. As such, we simulate all the single-qubit gates on the qubit from the start. Therefore, we can compute $\rho_{3,1}$ without the need for any measurement. Similarly, $\rho_{3,3}$ can be derived based on $\rho_{3,2}$ with a classically simulated Hadamard gate. As a result, we can eliminate the state preparation at $\rho_{3,2}$ and the measurement at $\rho_{3,3}$. This optimization also ensures the noiseless execution of the localized gates.

\textbf{Gate bypassing:}  Leveraging gate properties, we can reduce the required number of measurements. For example, the controlled gates do not change the Z basis distribution on the control qubit. Since we compute the density matrix $\rho_{3,1}$, there is no need to measure $\rho_{3,2}$ in the Z basis because the sequence of controlled U gates leaves the distribution unchanged. This sequence of gates can be bypassed when tracking the Z basis distribution, thereby ensuring noiseless computation on the Z basis.

\textbf{State traceback:} We can reduce the number of measurements and preparations when we trace back from the final measurements on the computational basis. Since only the measurement in the Z basis is necessary at the circuit's conclusion, we can omit the measurements on the X and Y bases. For instance, by following our discussion on gate bypassing, on the third qubit, the last two controlled-RZ gates do not alter the distribution on the Z basis. It suffices to acquire only the Z basis information of $\rho_{3,3}$. Since we simulate the single-qubit Hadamard gates, we only need to measure on the X basis for $\rho_{3,2}$ to acquire the Z basis distribution of $\rho_{3,3}$.

After the aforementioned optimizations, the circuit with the necessary preparations and measurements is shown in Fig.~\ref{fig:qpe_example}(c). Only four circuits are needed to calculate the output distribution on the third qubit. In the second step of the process, labeled \textbf{Circuit Execution and Error Mitigation}, we further implement the following optimizations:

\textbf{False dependency removal:} As we only need to measure a subset of qubits, we can remove the gates that the subset measurement is not dependent upon. This is similar to identifying causal cones in variational quantum ansatz studies~\cite{benedetti2021hardwareCausalCone, amaro2022filteringCausalcCone}. Nevertheless, there might be false dependencies in the circuit diagram. In the original circuit, the controlled-$U$ gate and controlled-$U^2$ gate may affect our measurement of $\rho_{3,2}$. By employing gate commuting rules, we can shift these gates after the measurement of $\rho_{3,2}$. Then, it becomes evident that the controlled-$U$ gate and controlled-$U^2$ gate can be removed. After this optimization, the gates that the $\rho_{3,2}$ measurement is dependent on are shown in Fig.~\ref{fig:qpe_example}(d).

\begin{figure*}
\centering
{\includegraphics[width=\linewidth]{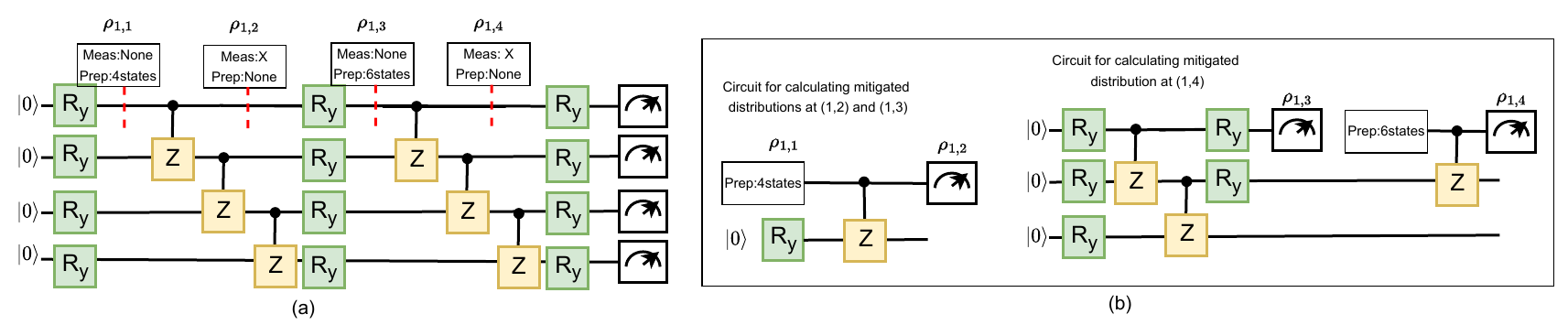}
}\hfil
\caption{QuTracer framework for two-layer VQE ansatz}
\label{fig:VQE_twolayer_example}
\end{figure*}

\textbf{Qubit remapping:}  Since the optimized circuit to be executed on the hardware is different from and smaller than the original circuit, we can remap it to physical qubits with low noise. The remapping ensures the low noise execution of the circuit. Furthermore, circuits for different subsets of qubits may re-utilize the same set of high-quality qubits.   
We employ the noise-aware mapping scheme~\cite{nation2023suppressingEnsemble} for our re-mapping.

\textbf{Qubit subsetting Pauli checks:} The final circuit shown in Fig.~\ref{fig:qpe_example}(d) can be protected by QSPC. As we prepare the input state in different states and acquire the measurement results from the real device, we can calculate the error mitigated distribution on the X basis following the discussion in Section~\ref{sec:qubit subsetting checks}. 
We start with the ground state $\ket{0}$ and calculate the noiseless density matrix $\rho_{3,1}$. Then, we run the circuit in Fig.~\ref{fig:qpe_example}(d) to acquire the noise-mitigated distribution on the X basis. Next, we simulate the Hadamard gate and bypass the Rz gates to get the final output distribution on the third qubit. In this process, the only step susceptible to hardware noise is the execution of the controlled gate $U^4$. The reduction of the circuit size, coupled with QSPC for error mitigation, makes the local distribution of the third qubit substantially more accurate than measurement subsetting. 

We can follow a similar procedure to obtain the accurate local distribution of the first and the second qubits. These qubits exhibit similar properties, and each only needs a single-qubit subsetting Pauli check.  Notably, this requirement remains consistent regardless of the size of the Quantum Phase Estimation (QPE) algorithm, where the single-qubit distribution only demands a single-qubit subsetting Pauli check per individual qubit. In the next section, we will use the multilayer VQE algorithm as an example to explore the challenges associated with sequentially implementing multiple qubit subsetting Pauli checks.
\subsection{Multi-layer Qubit Subsetting}

In this subsection, we illustrate the application of the QuTracer framework on a multi-layer ansatz circuit for variational quantum algorithms. As depicted in Fig.~\ref{fig:VQE_twolayer_example}(a), the ansatz circuit is composed of layers of single-qubit Y-rotation gates and linear entanglement with CZ gates. In our discussion, we concentrate on the top qubit. As illustrated in Fig.~\ref{fig:VQE_twolayer_example}(a), four circuit cuts are placed as two circuit cuts are needed per layer to ensure that the check is performed only on the qubit subset. For example, we can use pairs of Pauli Z checks at cut points (1,1) and (1,2) to protect one layer of CZ gates. However, if we use only the cut points (1,1) and (1,4) to check two layers, we cannot find pairs of single-qubit gates $C_L$ and $C_R$ that would satisfy the requirements for Pauli checks. The reason is when a single-qubit Pauli check is specified on one side, the corresponding check on the other side becomes a multi-qubit operation. Consequently, the operation would not be restricted to the subset, necessitating the tracking of additional qubits beyond the subset.

As we have identified the four cut points, we perform the aforementioned optimizations and construct the quantum circuits. As shown in Fig.~\ref{fig:VQE_twolayer_example}(b), each layer generates a circuit that can be protected with a single-qubit subsetting Pauli check. For the first circuit, the calculations yield the mitigated distributions at $(1,2)$ and $(1,3)$. However, for the second circuit, we measure the state at (1,3) and obtain an unmitigated density matrix $\rho_{1,3}$, which is then used to calculate the mitigated density matrix $\rho_{1,4}$. The challenge arises when considering the sequential execution of these circuits. If they run separately and in sequence, the mitigation is restricted to one layer only, with no provision to transmit the mitigated data to the subsequent calculation. Attempting to merge these two circuits to simultaneously check the two layers (akin to simultaneously cutting at all four cut points) would lead to an exponential increase in the required state preparation and measurement as the number of layers increases. The primary challenge lies in effectively transferring the mitigated distribution from one layer to the next.

The problem of transferring the mitigated distribution from one layer to the next can be translated into the task of updating the global output distribution based on the local output distributions. Referencing Fig.~\ref{fig:VQE_twolayer_example}(b), the circuit for calculating $\rho_{1,4}$ yields the output distribution $P = \{00:a, 01:b, 10:c, 11:d\}$ which is the measurement result for locations (1,3) and (1,4), and $P$ can be viewed as the global output distribution when measuring locations (1,3) and (1,4) simultaneously. Since we already obtain the error-mitigated output distribution at (1,3) represented as $M = \{0:\alpha, 1:\beta\}$ based on the calculation in the first layer, we can use $M$ to update the global output distribution $P$. The global output distribution is updated using the same Bayesian recombination algorithms utilized by SQEM~\cite{liu_2022classicalSimsAsQEMviaCircCut}. The updated global output distribution $P_{update}$ has a bitwise distribution for location (1,3) which is the same as the error-mitigated output distribution $M$. We can then employ the error-mitigated probability $P_{update}$ in the second layer's qubit subsetting calculation, which will mitigate the noise in both layers.

\subsection{\rev{Different Subset Sizes}}
\label{subsec:subset_size}
\rev{While increasing the subset size captures more global correlation, the noise also increases. The previous measurement subsetting works~\cite{Das_2021JigSawSubsetting, Dangwal_2023varsawAppTMeasEMForVQA} suggest that a subset size of 2 strikes a balance. Increasing the subset size allows us to ``virtualize" multiple Pauli checks concurrently. For instance, the VQE circuit shown in Fig.~\ref{fig:VQE_twolayer_example} consists of single-qubit Pauli Z checks. Extending the subset size to 2 allows simultaneous application of IZ and ZI checks, which detects more errors. However, as will be discussed in the following subsection, the classical and quantum overhead scales exponentially with the subset size. We restrict the subset size to one or two in our experiments. When running QuTracer with subset size of two, in the worst case scenario, it requires $30^2$ circuits. }

\rev{For circuits yielding symmetric output states, it's advisable to use a subset size greater than one. This is because measuring a single qubit typically produces a uniformly distributed bitwise outcome, which does not effectively refine the global distribution from the original circuit. For example, the $\mathbb{Z}_2$ symmetry in the MaxCut problem and the QAOA ansatz results in output states that are bit-flip invariant, with single qubit distributions being uniformly distributed. Therefore, the subset size should be larger than one. We will show the results for QAOA by setting the subset size to be two in Section~\ref{sec:evaluation}.}

\subsection{Scalability}
\label{subsec:scalability}

\rev{The cost of our proposed QuTracer framework includes classical circuit analysis, quantum circuit execution, and classical post-processing.} It scales linearly with the number of layers in the circuit. Consider the original circuit with $n$ qubits, $m$ layers, $k$ shots, and a qubit subset size of $s$. First, we consider the overhead in quantum processes. The total number of shots represents the total execution time on a quantum device. For each qubit subsetting Pauli check, we require $O(C^s)$ number of preparation and measurement circuits, whereas $C$ is the number of circuit copies for a single-qubit QSPC. Then, following the discussion in measurement subsetting~\cite{Das_2021JigSawSubsetting}, the number of subsets to be evaluated is $O(n)$. Since we have m layers, the total number of quantum circuits is $O(C^s nm)$. \rev{These circuits for the same check $O(C^s)$ need to be executed in sequential, while different checks $O(nm)$ can be parallelized. Since these quantum circuits only involve state preparation and measurement and do not rely on data from classical post-processing, they can be executed simultaneously with classical post-processing.} These quantum circuits require fewer shots than the original circuit. As we measure the expectation values, the number of shots~\cite{peruzzo2014VQE} scales polynomially with the number of qubits: $k\sim O(n^r/{\epsilon^2})$, where $r$ is a constant determined by the quantum circuit and is greater than one, and $\epsilon$ is the output error rate. Since we only measure a subset of qubits, the number of shots in each circuit copy $\epsilon$ is bounded by $O(\frac{s}{n}k)$ to maintain the same error rate. Therefore, the total number of shots in the QuTracer framework is $O(C^smk)$, which increases linearly with the number of layers in the circuit. As discussed in Section~\ref{sec:qubit subsetting checks}, if we limit the subset size to 1, the number of shots is upper-bounded by $O(30mk)$. As we will show in Section~\ref{subsec:realdevices}, the actual overheads are smaller due to our proposed optimizations. The constant $C$ can be further reduced by utilizing Local Operations and Classical Communication (LOCC), specifically through mid-circuit measurements and classically controlled operations. Following the discussion in~\cite{harada2023doubly}, we can prepare the states based on the measurement results and reduce the constant $C$ from 30 to 15. However, due to the lower fidelity associated with mid-circuit measurements and classically controlled operations, we did not implement these techniques in our experiments.

Next, we analyze the classical computation overhead. \rev{The circuit analysis simply traverses the circuit and has a complexity of $O(nm)$. The primary bottleneck in classical computation arises in post-processing the measured data to calculate the post-selected output states.} Since the number of basis states grows exponentially with the number of qubits, the classical computation overhead for each qubit subsetting check is $O(C'^s)$, where $C'$ is the time for processing the measurement results of all the circuit copies for a single-qubit QSPC. Since we limit the subset size, \rev{both the classical memory and computation overhead is $O(C'^s nm)$, which scales linearly with the number of qubits and the number of layers. Due to the data dependency across layers, classical computations for the same check on the same qubit ($O(C'^s m)$) are sequential. The computations for different qubits $O(n)$ can be performed in parallel. }

\section{Experimental Methodology}
\label{sec:methodology}
We evaluate our proposed QuTracer on noisy simulators and real quantum devices.

\textbf{Benchmarks:} The benchmarks in our experiments include Quantum Phase Estimation (QPE)~\cite{kitaev1995QPE}, Bernstein-Vazirani algorithm, the QFT adder~\cite{draper2000adder}, the QFT multiplier~\cite{ruiz2017quantum_multilexer}, Variational Quantum Eigensolver (VQE)~\cite{kandala2017hardwareVQE}, \rev{and Quantum Approximate Optimization Algorithm (QAOA)~\cite{farhi2014qaoa}}. 

\textbf{Implementation:} We implemented our QuTracer framework on top of the quantum computing framework Qiskit~\cite{Qiskit}.

\textbf{Devices:} We perform our noise simulation experiments on the Qiskit noisy simulator with noise models that incorporate single- and two-qubit gate errors and measurement errors.  
The real device experiments are conducted on the 27-qubit quantum device \texttt{ibm\_hanoi}, \rev{127-qubit quantum device \texttt{ibm\_kyoto}, and 127-qubit quantum device \texttt{ibm\_cusco}}.

\textbf{Evaluation metrics and setup:} We use Hellinger Fidelity~\cite{nikulin2001hellinger} as the evaluation metrics. Hellinger Fidelity serves as a measure of similarity between two probability distributions. We use it to evaluate the closeness between the noisy distributions to the ideal (i.e., noise-free) probability distributions.

We map and optimize the circuits with the maximum optimization level in Qiskit. Since the circuit transpilation is a stochastic process, we transpiled a circuit fifty times and selected the design with the least number of CNOTs. The number of shots for the original circuits running on the real device is set to 100000. The experiments for the same benchmark are executed in the same calibration cycle to avoid unexpected changes in hardware properties.

\textbf{Comparison with JigSaw and SQEM:} To evaluate the effectiveness of QuTracer, we compared it with the JigSaw and SQEM approach. JigSaw runs half of the shots in the global mode where all the qubits are measured, and the other half in the subset mode where only a subset of qubits are measured. Consistent with recommendations in prior measurement subsetting works~\cite{Dangwal_2023varsawAppTMeasEMForVQA, Das_2021JigSawSubsetting}, the subset size in JigSaw is configured to 2.

\section{Evaluation}
\label{sec:evaluation}

\subsection{Measurement Error Mitigation}
\label{subsec:exp_measurement_mitigation}
\begin{figure}[htbp]
\centerline{\includegraphics[width=\linewidth]{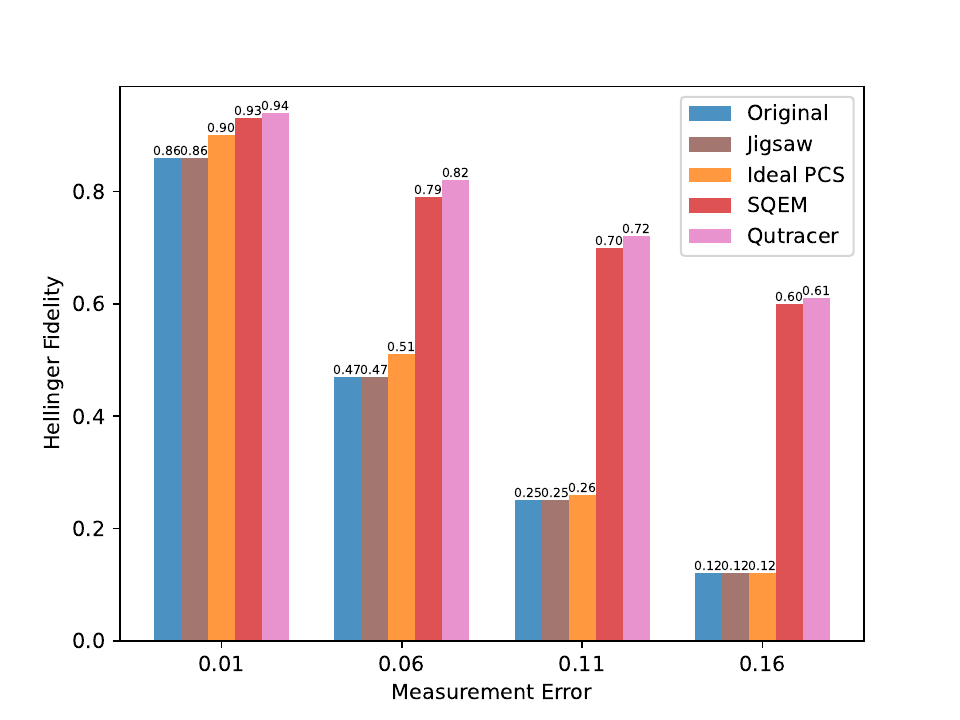}}
\caption{Change in Hellinger fidelity with respect to the measurement error.}
\label{fig:measurement_error} 
\end{figure}

In this experiment, we simulate a 15-qubit VQE circuit on Qiskit noisy simulator to study the measurement error mitigation effect of different error mitigation approaches.  
The 15-qubit VQE circuit has a similar circuit structure as shown in Fig.~\ref{fig:VQE_twolayer_example}(a), but with one layer of linear entanglement CZ gates. 
The subsetting size of QuTracer and SQEM approach is set to 1. We use a depolarization noise model in which each 1- and 2-qubit gate has depolarizing noise with a certain probability. The single- and two-qubit gate errors are fixed at 0.001 and 0.01, respectively. The noise model also incorporates \rev{uniform} single-qubit measurement errors. To study the effectiveness of the Pauli check circuits, we also simulate the circuits with the ideal Pauli checks (ideal PCS), i.e., the Pauli checks are implemented with no measurement errors on the ancilla qubits and no gate errors on the checking circuits. We use the Pauli Z checks shown in Fig.~\ref{fig:subset_check_example}(b). Fig.~\ref{fig:measurement_error} shows the output fidelities of different approaches as we vary the measurement error from 0.01 to 0.16. Since the noise model does not incorporate measurement crosstalk, the fidelity obtained from Jigsaw is similar to the fidelity of the original circuits. The circuits with ideal Pauli checks mitigate the gate errors and have shown improved fidelities over the original circuits. However, as we increase the measurement error, the measurement error becomes dominant and the fidelity improvement with ideal Pauli check circuits becomes less significant. Since SQEM and QuTracer both implement Pauli check circuits virtually, they are capable of mitigating the measurement noise. As the measurement error increases, both approaches mitigate a large proportion of the measurement error and lead to significant fidelity improvements. QuTracer achieves a higher fidelity than SQEM due to its optimizations and the reduced number of basis measurements. Based on the experiments, we evidently show that the ``virtual" implementation of the Pauli checks enables the mitigation of measurement errors.

\subsection{Gate Error Mitigation}
\begin{figure}[htbp]
\centerline{\includegraphics[width=\linewidth]{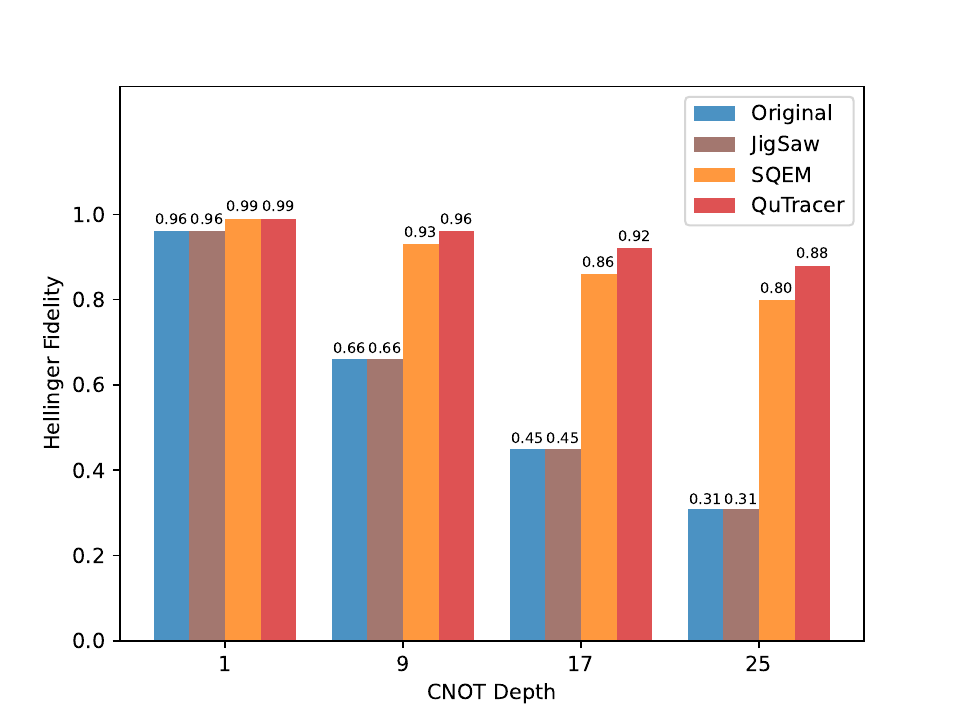}}
\caption{Changes in Hellinger fidelity with respect to the CNOT depth.}
\label{fig:gate_error} 
\end{figure}

In this experiment, we conduct simulations of an 8-qubit VQE circuit using Qiskit noisy simulator to study the gate error mitigation effect of different error mitigation approaches. 
The subsetting size of QuTracer and SQEM approach is set to 1. We use a depolarization noise model with a single-qubit gate error of 0.001, a two-qubit gate error of 0.01, and a single-qubit measurement error of 0.001 for all qubits. In order to assess the influence of gate errors, we vary the CNOT depth in the circuit by changing the repetition times of the linear entanglement gate layer, ranging from 1 to 25. Fig.~\ref{fig:gate_error}  shows the Hellinger fidelity under different error mitigation methods. JigSaw shows the same fidelity as the original circuit, as it does not mitigate gate errors. In comparison, both SQEM and QuTracer exhibit fidelity improvement. With increasing circuit depth, the fidelity gap between SQEM and QuTracer widens. This discrepancy arises because higher circuit depths introduce more gate errors. SQEM necessitates a state reconstruction process involving the preparation and measurement on all bases using the original circuit. As more gate noise is introduced, the accuracy of the state reconstruction process diminishes. On the other hand, QuTracer utilizes the optimization of False Dependency Removal to eliminate unnecessary gates, allowing it to outperform SQEM. A gate count comparison in Section~\ref{subsec:realdevices} confirms this distinction.

\label{subsec:exp_gate_error}
\subsection{\rev{Multi-layer Qubit Subsetting}}
\label{subsec: multi-layer}

\rev{In this experiment, we use a 10-qubit QAOA circuit on MaxCut problems with four layers to study the effectiveness of multi-layer qubit subsetting in a realistic noise model. The subsetting size of QuTracer is set to 2. 
The noise model and coupling map are from the quantum device \texttt{ibmq\_mumbai}. 
This noise model incorporates various factors, including gate errors, gate time, $T_1$ and $T_2$ relaxation times, and readout errors for each qubit. The parameter for this noise model is based on the calibration data collected on February 22, 2024. The median CNOT error is  $7.611\times10^{-3}$, the median gate time is $426.667\,\text{n}\text{s}$, the median readout error is $1.810\times10^{-2}$, the median T1 is $125.94\,\mu\text{s}$, and the median T2 is $188.75\,\mu\text{s}$.}

\begin{figure}[htbp]
\centerline{\includegraphics[width=\linewidth]{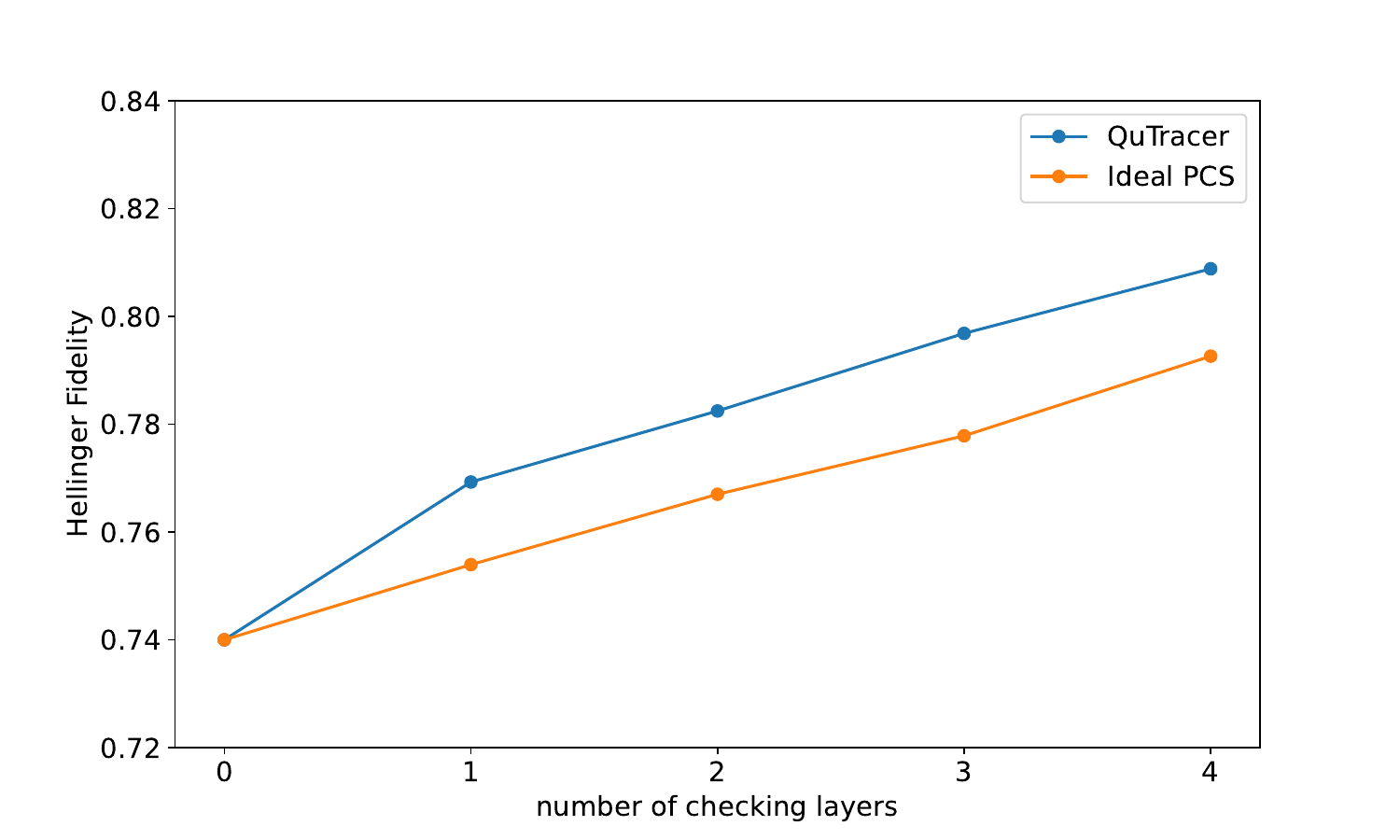}}
\caption{Changes in Hellinger fidelity with respect to the number of checking layers.}
\label{fig:fidelity} 
\end{figure}

\rev{The results are shown in Fig.~\ref{fig:fidelity}. We observe that as the number of checked layers increases, the fidelity improvement from QuTracer becomes more significant. Checking only the fourth layer improves the fidelity by $3.96\%$, checking both the third and the fourth layer improves the fidelity by $5.74\%$, checking the second, the third, and the fourth layer improves the fidelity by $7.68\%$, and checking all layers improves the fidelity by $9.42\%$, showing that sequentially checking multiple layers is effective in error mitigation.}

\rev{We also compare with ideal Pauli checks (ideal PCS), i.e., the Pauli checks are implemented with no measurement errors on the ancilla qubits and no gate errors on the checking circuit. 
Interestingly, QuTracer surpasses the performance of the ideal PCS.
This is because when QuTracer checks multiple layers, it can take advantage of utilizing the optimization of False Dependency Removal to remove unnecessary gates in the circuits for each of the layers, while PCS checks multiple layers simultaneously and can not take advantage of the optimizations for each of the layers, which results in a larger circuit with higher noise.
}

\begin{table*}[htbp]
\centering
\caption{\rev{Simulation results for QAOA circuits with different layers.}}
\label{table:simulation_results_QAOA_vary_iteration}
\resizebox{\linewidth}{!}{%
\begin{tabular}{|c|c|c|c|c|c|c|c|c|c|c|}
\hline
&  \multicolumn{3}{c|}{Normalized number of shots} &  \multicolumn{3}{c|}{Average 2-qubit basis gate count} &  \multicolumn{3}{c|}{Hellinger Fidelity} & \multicolumn{1}{c|}{Fidelity Improvement}\\ \hline
Workload  & Original & JigSaw &  QuTracer & Original & JigSaw &  QuTracer   & Original & JigSaw &  QuTracer &   QuTracer\\ \hline
10-q QAOA with 1 layers  &1 &1 &16 &26 &26 &6 &0.90 &0.90 &0.92 &2.89\%\\ \hline
10-q QAOA with 2 layers  &1 &1 &106 &52 &52 &21 &0.80 &0.80 &0.83 &3.58\% \\ \hline
10-q QAOA with 3 layers  &1 &1 &196 &78 &78 &29 &0.78 &0.79 &0.84  &8.41\%\\ \hline
10-q QAOA with 4 layers  &1 &1 &286 &104 &104 &37 &0.74 &0.74 &0.81 &9.42\%\\ \hline
10-q QAOA with 5 layers  &1 &1 &376 &130 &130 &47 &0.59 &0.60 &0.70 &18.09\%\\ \hline
\end{tabular}
}
\end{table*}
\begin{table*}[htbp]
\centering
\caption{Real device results for single-layer circuits. }
\label{table:real_device_single_layer}
\resizebox{\linewidth}{!}{%
\begin{tabular}{|c|c|c|c|c|c|c|c|c|c|c|c|c|}
\hline
  &  \multicolumn{4}{c|}{Normalized number of shots} &  \multicolumn{4}{c|}{Average 2-qubit basis gate count} &  \multicolumn{4}{c|}{Hellinger Fidelity} \\ \hline
Workload  & Original & JigSaw & SQEM & QuTracer & Original & JigSaw & SQEM &  QuTracer  & Original & JigSaw & SQEM &  QuTracer\\ \hline
4-q QFTMultiplier   &1 &1 &N/A &11 &28 &28 &N/A &18 &0.49 &0.49 &N/A &0.65 \\ \hline
5-q QPE             &1 &1 &N/A &11 &29 &29 &N/A &11 &0.20 &0.20 &N/A &0.49 \\ \hline
6-q QPE             &1 &1 &N/A &11 &44 &44 &N/A &17 &0.19 &0.19 &N/A &0.29 \\ \hline
7-q QFTAdder        &1 &1 &N/A &15 &75 &75 &N/A &37 &0.22 &0.22 &N/A &0.35 \\ \hline
9-q BV              &1 &1 &13 &11 &21 &21 &21 &2 &0.07 &0.09 &0.13 &0.89 \\ \hline
12-q VQE with 1 layer           &1 &1 &13 &11 &11 &11 &11 &2 &0.67 &0.76 &0.88 &0.96 \\ \hline
15-q VQE with 1 layer          &1 &1 &13 &11 &14 &14 &14 &2 &0.36 &0.50 &0.65 &0.87 \\ \hline
\rev{10-q QAOA with 1 layer}          &1 &1 &N/A &16 &26 &26 &N/A &6 &0.57 &0.57 &N/A &0.86 \\ \hline
\end{tabular}
}
\end{table*}

\begin{table*}[htbp]
\centering
\caption{Real device results for circuits with multiple layers.}
\label{table:real_device_multi_layer}
\resizebox{\linewidth}{!}{%
\begin{tabular}{|c|c|c|c|c|c|c|c|c|c|}
\hline
&  \multicolumn{3}{c|}{Normalized number of shots} &  \multicolumn{3}{c|}{Average 2-qubit basis gate count} &  \multicolumn{3}{c|}{Hellinger Fidelity} \\ \hline
Workload  & Original & JigSaw &  QuTracer & Original & JigSaw &  QuTracer   & Original & JigSaw &  QuTracer\\ \hline
12-q VQE with 2 layers  &1 &1 &29 &22 &22 &7 &0.37 &0.52 &0.65  \\ \hline
12-q VQE with 3 layers  &1 &1 &47 &33 &33 &10 &0.29 &0.39 &0.49 \\ \hline
15-q VQE with 2 layers  &1 &1 &29 &28 &28 &7 &0.21 &0.28 &0.69  \\ \hline
15-q VQE with 3 layers  &1 &1 &47 &42 &42 &11 &0.06 &0.06 &0.54 \\ \hline
\rev{10-q QAOA with 2 layers}  &1 &1 &106 &52 &52 &21 &0.16 &0.28 &0.36 \\ \hline
\rev{10-q QAOA with 3 layers}  &1 &1 &196 &78 &78 &29 &0.14 &0.16 &0.40 \\ \hline
\end{tabular}
}
\end{table*}

\subsection{\rev{Scaling of Circuit Depth}}
\label{subsec:depthscaling}
\rev{
In this subsection, we use 10-qubit QAOA circuits on MaxCut problems with layers ranging from 1 to 5 to evaluate QuTracer's efficiency in handling circuits of varying depths in a realistic noise model. The detailed parameters of the noise model are discussed in Section~\ref{subsec: multi-layer}. The subsetting size of QuTracer is set to 2. The results are shown in Table~\ref{table:simulation_results_QAOA_vary_iteration}.}

\rev{For the QAOA circuits, QuTracer interprets each circuit step as a distinct layer, applying Multi-layer Qubit Subsetting to check each of the layers.
In examining individual layers, the necessity to check multiple pairs of qubits can be reduced by the nature of the problem addressed by QAOA. Specifically, when the application of QAOA is to solve MaxCut problems on regular graphs, the QAOA circuits exhibit symmetric properties. As a result, checking one pair of qubits can yield results for multiple qubit pairs located in symmetric positions.
}

\rev{Results presented in Table~\ref{table:simulation_results_QAOA_vary_iteration} indicate that as circuit depth increases, the Hellinger Fidelity of the original circuits, prior to the application of any error mitigation strategies, diminishes due to the introduction of additional noisy layers. However, the efficacy of QuTracer in enhancing Hellinger Fidelity becomes more significant with deeper circuits, Table~\ref{table:simulation_results_QAOA_vary_iteration} includes the fidelity improvement of QuTracer over the unmitigated results, showing QuTracer's capability to effectively mitigate errors in deeper circuits.}

\rev{We also compared with JigSaw, which shows little improvement in Hellinger Fidelity. This is due to the Qiskit device noise model, which does not account for cross-talk noise, thereby impacting JigSaw's performance.
}
\subsection{Real Device Experiments}
\label{subsec:realdevices}

The experimental results on the real device are shown in Tables~\ref{table:real_device_single_layer} and~\ref{table:real_device_multi_layer}. Applying SQEM to the benchmarks QFTMultiplier, QPE, QFTAdder, and QAOA is not practical as it incurs an exponential overhead. Therefore, we mark parts of the results in SQEM as N/A. The experiments on the same benchmark are executed in the same calibration cycle, while those for different benchmarks may be performed in different cycles with different hardware properties.

We initiate our discussion by focusing on the benchmarks that require only a single-layer QSPC in QuTracer. As shown in Table~\ref{table:real_device_single_layer}, the benchmarks include QFTMultiplier, QPE, QFTAdder, VQE \rev{and QAOA} with a single layer of CZ gates. \rev{ QFTMultiplier, QPE, QFTAdder, and VQE are running on the 27 qubit machine \texttt{ibm\_hanoi}, the subsetting size of QuTracer is set to 1; QAOA is running on 127 qubit macine \texttt{ibm\_kyoto}, the subsetting size of QuTracer is set to 2.} The results with QuTracer show the highest output fidelity with an average of $2.3\times$ , $2.03\times$ and $2.15\times$ improvement over the unmitigated results, the results with JigSaw, and SQEM. In the VQE example, when increasing the number of qubits from 12 to 15, JigSaw's results exhibit a noticeable decrease from $0.76$ to $0.50$.  As shown in the table, the average CNOT counts for the QuTracer framework are significantly smaller than other approaches, which leads to improved output fidelity.

 Table~\ref{table:real_device_multi_layer} shows the results of different schemes on the VQE algorithm \rev{and QAOA} with multiple layers of CZ gates. \rev{VQE benchmark is running on 27 qubit machine \texttt{ibm\_hanoi}, the subsetting size of QuTracer is set to 1; QAOA benchmark is running on a 127-qubit machine \texttt{ibm\_cusco}, the subsetting size of QuTracer is set to 2.} The SQEM framework is not included as its overhead scales exponentially with the number of layers. QuTracer improves the output fidelity by up to $9\times$ ($3.06\times$ on average) compared to the original circuit and by up to $9\times$ ($2.43\times$ on average) compared to JigSaw. As we increase the number of layers in the circuit, gate errors begin to play a more significant role. The superior fidelity improvement of QuTracer comes from its ability to effectively mitigate the increased gate errors.

\section{Discussion}
\subsection{Integration with Other Error Mitigation Techniques}
Our proposed QuTracer framework is complementary with other error mitigation techniques. QuTracer generates multiple copies of the original circuit with fewer gates and fewer qubits. Existing error mitigation techniques such as dynamical decoupling~\cite{das2021dynamical_decoupling}, zero-noise extrapolation~\cite{Giurgica_2020DigitZNEFroQEM},  Clifford data regression~\cite{Czarnik_2021errormitigationWithCliffQCData}, and probabilistic error cancellation~\cite{berg2022probabilistic_scalable_mitigation} can be applied to improve the fidelity of circuit copies on hardware. Moreover, the reduced size of the circuit copies enables a better scaling of mitigation approaches. For example, Clifford data regression requires fewer training data for smaller circuits. The study of integrating QuTracer with other error mitigation techniques is left for future work.
\subsection{Application-tailored Extension of QuTracer}
Prior work~\cite{Dangwal_2023varsawAppTMeasEMForVQA} presents a measurement subsetting framework, Varsaw, that is designed to mitigate measurement noise in variational quantum algorithms (VQA). Compared to Jigsaw, the Varsaw framework reduces the computational cost by identifying spatial redundancy across subsets from different VQA circuits and temporal redundancy across global distributions from different VQA iterations. The optimizations in Varsaw can be seamlessly integrated into the QuTracer framework to produce an application-specific extension that reduces computational overhead. 
\section{Conclusion}
\label{sec:conclusion}
In this paper, we propose a qubit subsetting framework, QuTracer, to mitigate both gate and measurement errors. The key idea is to track a subset of qubits to mitigate the error along the computation process so as to achieve high-fidelity local distributions, which are then used to refine the global distribution. 
Our experimental results on noisy simulators and real quantum devices show that the proposed framework significantly outperforms current state-of-the-art approaches.

\section*{Acknowledgements}
We thank the anonymous reviewers for their valuable comments. The work is funded in part by NSF grants 1818914, 2325080 (with a subcontract to NC State University from Duke University), NSF grant 2120757 (with a subcontract to NC State University from the University of Maryland), and the U.S. Department of Energy, Office of Science, National Quantum Information Science Research Centers.


\bibliographystyle{IEEEtranS}
\bibliography{refs}
\newpage
\appendix
\section{Artifact Appendix}

\subsection{Abstract}

Our artifact provides the source code for QuTracer, and the Python scripts to run benchmarks on simulators and real quantum machines. These scripts also generate results for the evaluation part. Users can reproduce the results in Table~\ref{table:simulation_results_QAOA_vary_iteration}, ~\ref{table:real_device_single_layer} and~\ref{table:real_device_multi_layer}.

\subsection{Artifact check-list (meta-information)}

{\small
\begin{itemize}
  \item {\bf Compilation: } Qiskit transpiler
  \item {\bf Data set: }Benchmarks listed in Section~\ref{sec:methodology} 
  \item {\bf Run-time environment: }Ubuntu 20.04.6 LTS
  \item {\bf Hardware: }The experiments on real quantum machines require access to IBM quantum machines
  \item {\bf Execution: } Run the bash scripts and python scripts
  \item {\bf Metrics: }Hellinger fidelity, 2-qubit basis gate count, normalized number of shots
  \item {\bf Output: }JSON files and CSV files
  \item {\bf Experiments: }Apply different Error Mitigation methods to quantum circuits and compare the normalized number of shots, 2-qubit basis gate count, and Hellinger fidelity.
  \item {\bf How much disk space required (approximately)?: }2GB
  \item {\bf How much time is needed to prepare workflow (approximately)?: }1 hour
  \item {\bf How much time is needed to complete experiments (approximately)?: }12+ hours
  \item {\bf Publicly available?: }Yes
  \item {\bf Code licenses (if publicly available)?: }Apache-2.0 License
  \item {\bf Archived (provide DOI)?: } 10.5281/zenodo.11075556
\end{itemize}
}

\subsection{Description}

\subsubsection{How to access}

The source code for QuTracer and the scripts to run the benchmarks are available in github: \url{https://github.com/peiyi1/QuTracer_project}
\subsubsection{Hardware dependencies}
The access to IBM quantum machines is needed to reproduce the results in Table~\ref{table:real_device_single_layer} and~\ref{table:real_device_multi_layer}.  
\subsubsection{Software dependencies}
Python 3.10 is used in our experiments.
\subsubsection{Data sets}
Benchmarks are listed in Section~\ref{sec:methodology} 

\subsection{Installation}

\subsubsection{Create conda environment}
Download Anaconda at https://www.anaconda.com/ and install it.
\vspace{\baselineskip}

Create an environment named qutracer:

\$ conda create -y -n qutracer python=3.10
\vspace{\baselineskip}

Activate the environment:

\$ conda activate qutracer 

\vspace{\baselineskip}

\subsubsection{Install Qiskit and other necessary packages}
The version of Qiskit used in our experiment is 0.45.1.
\vspace{\baselineskip}

\$ pip install qiskit==0.45.1

\$ pip install qiskit-ibm-provider==0.7.3

\$ pip install qiskit-aer==0.13.2

\$ pip install retworkx==0.13.2

\$ pip install networkx==3.2.1

\$ pip install matplotlib

\$ pip install pylatexenc

\$ pip install pyyaml

\vspace{\baselineskip}

\subsubsection{QuTracer package installation}

Clone the repository of QuTracer:

\$ git clone https://github.com/peiyi1/QuTracer\_project.git

\vspace{\baselineskip}

Go to the path /QuTracer\_project/QuTracer and install QuTracer:

\$ cd QuTracer\_project/QuTracer

\$ pip install .

\subsection{Experiment workflow}
\subsubsection{experiments on Qiskit AerSimulator}
In the path /QuTracer\_project/test\_on\_simulator, there are three directories: script, yaml\_file, and saved\_data. The directory script contains all the scripts that can generate circuits, execute circuits, and process the results for the circuits in different error mitigation methods. The directory yaml\_file contains all the configuration files for different benchmarks, users can modify the configuration file to set up different configurations for running the benchmark. The directory saved\_data contains all the saved circuits and results in JSON files. To reproduce the results in the directory saved\_data, run the following command.
\vspace{\baselineskip}

Go to the path /QuTracer\_project/test\_on\_simulator:

\$ cd QuTracer\_project/test\_on\_simulator/
\vspace{\baselineskip}

Generate all the circuits needed in different error mitigation methods:

\$ ./generate\_circuits.sh
\vspace{\baselineskip}

Execute all the circuits generated in the previous step:

\$ ./execute\_circuits.sh
\vspace{\baselineskip}

Run the following command to process the results from all the circuits and calculate the Hellinger fidelity for different error mitigation methods. All the results are saved in the directory saved\_data. For example, the results for 10-qubit QAOA with 1 layer are saved in the path /QuTracer\_project/test\_on\_simulator/saved\_data/ qaoa\_reps1\_n10/results

\$ ./results\_postprocessing.sh 
\vspace{\baselineskip}

\subsubsection{experiments on IBM quantum machine}\label{subsubsec:workflow_real_machine}
In the path /QuTracer\_project/test\_on\_real\_machine, there are three directories: script, yaml\_file, and saved\_data. For all the YAML files in the directory yaml\_file, the configuration option 'enable\_running\_on\_real\_machine' is set to 'True' to run all the circuits on real IBM quantum machines. To reproduce the results in the directory saved\_data, run the following command.
\vspace{\baselineskip}

Go to the path /QuTracer\_project/test\_on\_real\_machine:

\$ cd QuTracer\_project/test\_on\_real\_machine/
\vspace{\baselineskip}

Generate all the circuits needed in different error mitigation methods:

\$ ./generate\_circuits.sh
\vspace{\baselineskip}

Execute all the circuits generated in the previous step:

\$ ./execute\_circuits.sh
\vspace{\baselineskip}

After running the above command, all the circuits have been sent to the job queue of IBM quantum machines, the job status can be checked on the IBM Quantum Platform website: https://quantum.ibm.com/jobs. Wait until all the job statuses become completed, then run the following command to load and save all the job results:

\$ ./save\_results\_real\_machine.sh
\vspace{\baselineskip}

Process the results from all the circuits and calculate the Hellinger fidelity for different error mitigation methods:

\$ ./results\_postprocessing.sh 

\subsection{Evaluation and expected results}
After going through all the steps in the Experiment Workflow section, go back to the path /QuTracer\_project and run following commands to generate CSV files, which correspond to the results in Table~\ref{table:simulation_results_QAOA_vary_iteration}, ~\ref{table:real_device_single_layer} and~\ref{table:real_device_multi_layer}.
\vspace{\baselineskip}

Generate the file simulation\_results\_for\_qaoa.csv, which contains results in Table~\ref{table:simulation_results_QAOA_vary_iteration}:

\$ python generate\_table\_simulation\_results\_for\_qaoa.py 
\vspace{\baselineskip}

Generate the file real\_machine\_results\_for\_single\_layer\_

circuits.csv, which contains results in Table~\ref{table:real_device_single_layer}:

\$ python generate\_table\_real\_machine\_results\_for\_single\_

layer\_circuits.py
\vspace{\baselineskip}

Generate the file real\_machine\_results\_for\_circuits\_with\_

multiple\_layers.csv, which contains results in Table~\ref{table:real_device_multi_layer}:

\$ python generate\_table\_real\_machine\_results\_for\_

circuits\_with\_multiple\_layers.py

\subsection{Experiment customization}

The configuration in the YAML files can be modified to run benchmarks with different configurations.
\subsection{Notes}
The Experiment Workflow subsection~\ref{subsubsec:workflow_real_machine} requires access to IBM quantum machines, and the waiting time for running circuits on IBM quantum machines may vary based on the size of the waiting queue.  

\subsection{Methodology}

Submission, reviewing and badging methodology:

\begin{itemize}
  \item \url{https://www.acm.org/publications/policies/artifact-review-and-badging-current}
  \item \url{http://cTuning.org/ae/submission-20201122.html}
  \item \url{http://cTuning.org/ae/reviewing-20201122.html}
\end{itemize}

\end{document}